\documentclass[usegraphicx]{mn2e}

\usepackage{fixltx2e}
\usepackage{mathptmx}

\include{defn}
\begin{document}

\title[Reflection on truncation and spin ]{On the determination of the
  spin and disc truncation of accreting black holes using X-ray
  reflection} \author[A.C. Fabian et al] {\parbox[]{6.5in}{{
      A.C.~Fabian$^1\thanks{E-mail: acf@ast.cam.ac.uk}$,
      M.L.~Parker$^1$, D.R.~Wilkins$^1$, J.M.~Miller$^2$, E.~Kara$^1$,
      C.S.~Reynolds$^3$ and T. Dauser$^4$
    }\\
    \footnotesize
    $^1$ Institute of Astronomy, Madingley Road, Cambridge CB3 0HA\\
    $^2$ Department of Astronomy, University of Michigan, Ann Arbor,
    MI 48109, USA\\
    $^3$ Department of Astronomy, University of Maryland, College
    Park, MD 20742-2421, USA\\
$^4$ Dr Karl Remeis-Observatory and Erlangen Centre for Astroparticle
Physics, Sternwartstr. 7, D-96049 Bamberg, Germany\\
  }}
\maketitle
  
\begin{abstract}
  We discuss the application of simple relativistically-blurred X-ray
  reflection models to the determination of the spin and the inner
  radius of the disc in accreting black holes. Observationally, the
  nature of the corona is uncertain {\it a priori}, but a robust
  determination of the inner disk radius can be made when the disc
  emissivity index is tightly constrained.  When the inner disc is
  well illuminated, the black hole spin can also be determined.  Using
  reflection modelling derived from ray tracing, we show that robust
  determination of disc truncation requires that the location of the
  coronal source is quasi-static and at a height and radius less than
  the truncation radius of the disc. Robust spin measurements require
  that at least part of the corona lies less than about 10
  gravitational radii above the black hole in order that the innermost
  regions, including the innermost stable circular orbit, are well
  illuminated.  The width of the blurring kernel (e.g., the iron line)
  has a strong dependence on coronal height.  These limitations may be
  particularly applicable at low Eddington fractions (e.g. the
  low/hard state, and low-luminosity AGN) where the height of the
  corona may be relatively large, or outflowing, and tied to jet
  production.
\end{abstract}


\section{Introduction}

X-ray reflection is becoming a common tool in X-ray astronomy,
particularly for studying the inner regions around compact objects. It
occurs when matter is irradiated by a primary X-ray source, giving
rise to backscattering, fluorescence and secondary\footnote{Primary
  here means that the energy derives directly from the accretion
  process, through magnetic fields for example, whereas secondary
  means that the energy source is the irradiating X-rays.} emission
(see Fabian \& Ross 2010 for a review). The primary source is here
referred to as the corona. We initially assume it to be pointlike,
static and on axis, but in practice it may be extended and outflowing,
perhaps the base of a jet. The reflection spectrum provides
diagnostics for elemental abundances and ionization parameters of the
matter as well as its velocity and radius range through doppler and
gravitational redshifts. For sources with significant variability,
reverberation of the reflection component compared with the primary
can reveal the source size and geometry in physical units. Here we are
concerned with the use of X-ray reflection to measure the innermost
radius of the accretion disc around a black hole. If this is
identified with the Innermost Stable Circular Orbit (ISCO) then the
spin of the black hole can be determined. Alternatively, the disc may
be truncated well before the ISCO in which case the truncation radius
may be located.

Considerable effort has been expended recently on tracing light rays
in the Kerr metric in order to determine the emissivity profile on the
disc plane produced by a source above the disc (Wilkins \& Fabian
2012; Dauser et al 2013). The emissivity profile and the radius range
over which it is relevant are key factors in constructing the spectral
response of the reflection spectrum. Here we discuss limitations to
the measurement of spin and disc truncation when the primary source or
corona is much more than 10 gravitational radii ($10GM/c^2$) above the
disc. When that occurs, the innermost regions around the ISCO are
poorly irradiated so making the reflection spectrum from that part of
the disc, which is crucial to spin determination, difficult to
measure. This issue has been discussed before (e.g. Vaughan \& Fabian
2004; Chiang et al 2012; Dauser et al 2013) but is not widely
recognised. The purpose of this paper is to explore it further.

Spin determinations of AGN using X-ray reflection have been ongoing
since Dabrowski et al (1997) with recent results including
MCG--6-30-15 ($a=cJ/ GM^2=0.989^{+0.009}_{-0.002}$, Brenneman \&
Reynolds 2006), 1H0707-495 ($a>0.98$, Fabian et al 2009);
IRAS13224-3809 ($a=0.989\pm0.001$, Fabian et al 2013a) (see recent
reviews by Reynolds 2013 and Brenneman 2013 for summaries of more AGN
spin results). The high statistical precision of the above three
results is due to a combination of low coronal height and high iron
abundance, which emphasizes features in the reflection spectrum. X-ray
reflection-based spin determinations in black hole binaries (BHB) are
reviewed by Miller (2007) following Miller et al (2002, 2004); more
recent results include Cyg X-1 ($a=0.97^{+0.014}_{-0.02}$, Fabian et
al 2012; $a=0.98\pm0.01$, Tomsick et al 2014) and GRS~1915+105
($a=0.98\pm0.01$, Miller et al 2013). Objects with intermediate spin,
include SWIFT J2127.4+5654 ($a=0.6\pm0.2$, Miniutti et al 2009), XTE J
1752-223 ($a=0.52\pm0.11$, Reis et al 2011) and Fairall~9
($a=0.52^{+0.19}_{-0.15}$, Lohfink et al 2012). Complementary work on
BHB using the thermal continuum method (McClintock, Narayan \& Steiner
2013) is providing general agreement with X-ray reflection methods,
for objects where both methods can be applied (e.g. Steiner et al
2012).

Both methods rely on the identification of $r_{\rm in}$, the innermost
disc radius detected, with the ISCO\footnote{In principle both $r_{\rm
    in}$ and $a$ can be measured, but the deviations to the energy
  shifts produced at a given radius by spin alone are small, see e,g,
  Fig.~2 of Dovciak et al (2004)}. This systematic uncertainty is being
tackled with numerical simulations of discs (Reynolds \& Fabian 2008;
Shafee et al 2008; Noble et al 2011; Penna et al 2010; Kulkarni et al
2011). The simulations indicate that $r_{\rm in}$, the inner radius of
the dense disc (which is relevant to reflection), lies within a
pressure scale-height of the ISCO, which corresponds to an uncertainty
of $\sim (L/L_{\rm Edd})r_{\rm g}$.  Different methods may give
(slightly) different values for $r_{\rm in}$ depending upon their
sensitivity to surface density or emission. The disc may be thicker
when the thermal continuum method is optimally applied than for the
reflection approach. The low ionization parameter often found with the
reflection method for the inner disc of AGN (e.g. $\xi < 20$ for
1H0707-495 and IRAS13224-3809) indicates that the disc is dense and
thus probably thin. There are currently no published simulations of
the ISCO region of thin irradiated discs.

Disc truncation in which the optically-thick, physically-thin, disc
terminates beyond the ISCO and the inner region is filled with a hot
advection-dominated flow is often invoked as a mechanism for
explaining state changes in stellar mass black hole sources (Esin et
al 1997; Done et al 2007). In this model a source in the low, hard
state is interpreted as having a truncated disc. Spectral evidence for
truncation is confused in the literature with some work claiming that
it does not occur (e.g. Miller et al 2004, 2006, Reis et al 2008, 2010)
and some (Done \& Diaz Trigo 2010) that it does. Recent work by Plant
et al (2013) on the source GX339-4, which has featured much in this
debate, appears to show disc truncation in the low state using a
simple relativistic reflection model (see also Petrucci et al
2013). As discussed here in Section 2, this interpretation may be
too simplistic. Truncation has also been inferred in some radio-loud AGN, 
such as 3C120 (Marscher et al 2003; Lohfink et al 2013).

\section{Determination of the inner radius of the disc by reflection}

\begin{figure}
  \centering
  \includegraphics[width=0.99\columnwidth,angle=0]{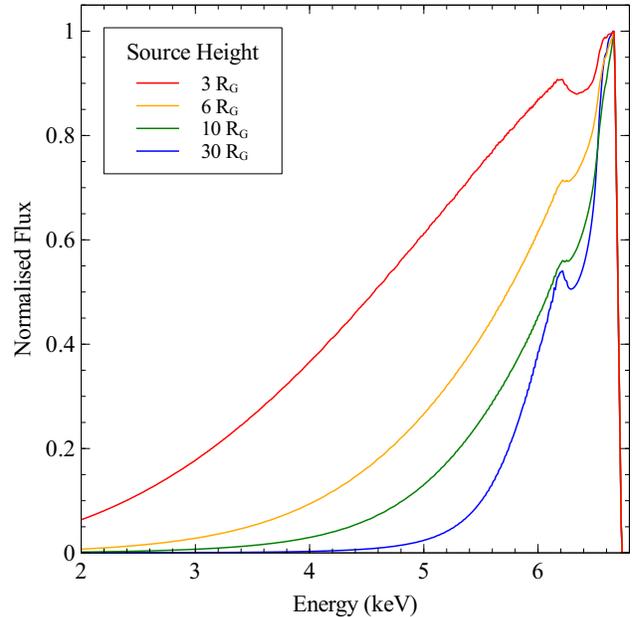}
  \caption{Model iron line profiles expected from an on-axis point source
    at heights of 3, 6, 10 and $30 r_{\rm g}$ above an accretion disc
    inclined at 30 deg around a rapidly spinning black hole with spin
    $a=0.998$. The profiles were obtained using
    {\sc relline-lp} of Dauser et al 2013.    }
\end{figure}

\begin{figure}
  \centering
  \includegraphics[width=0.99\columnwidth,angle=0]{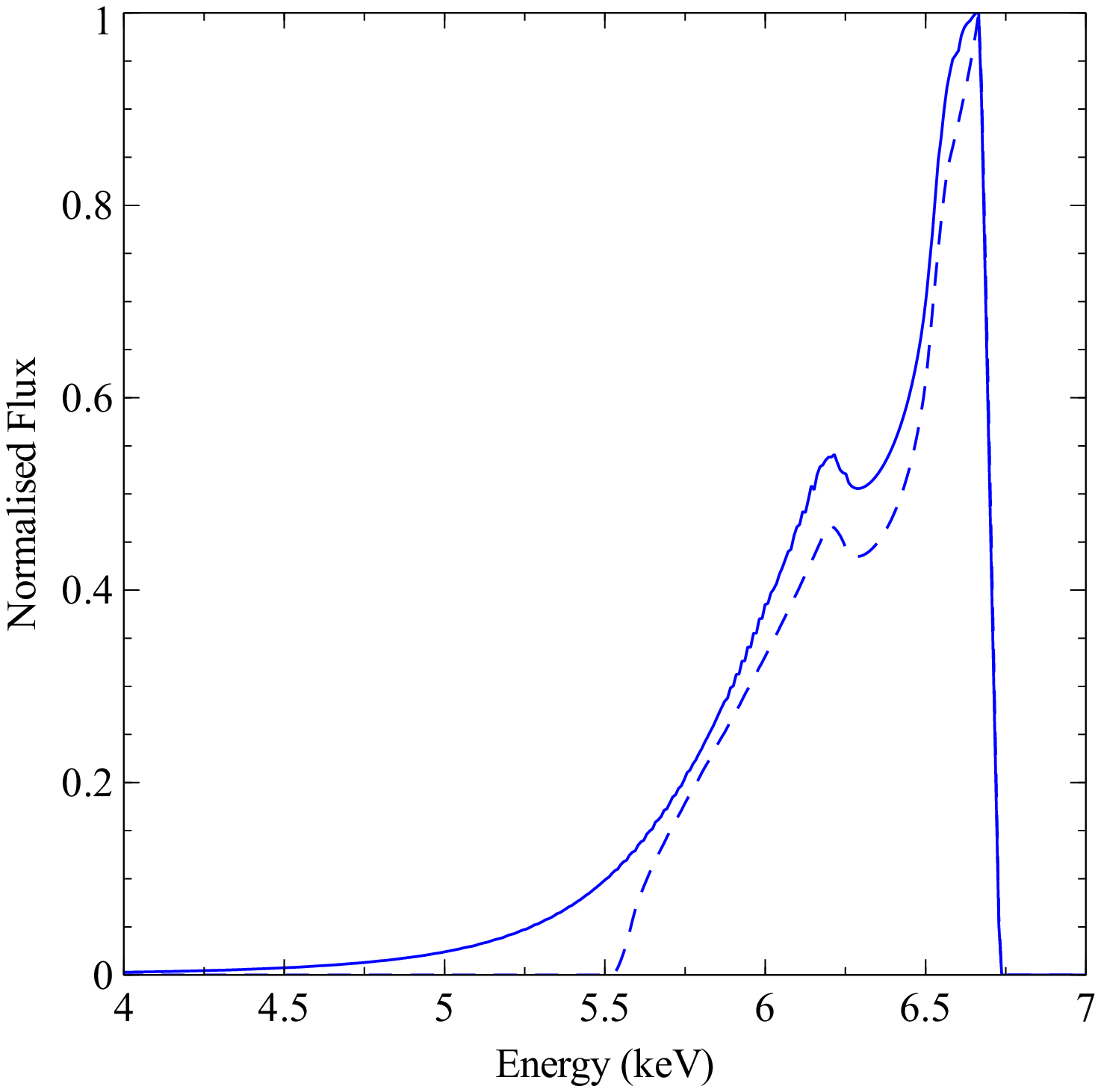}  
  \includegraphics[width=0.99\columnwidth,angle=0]{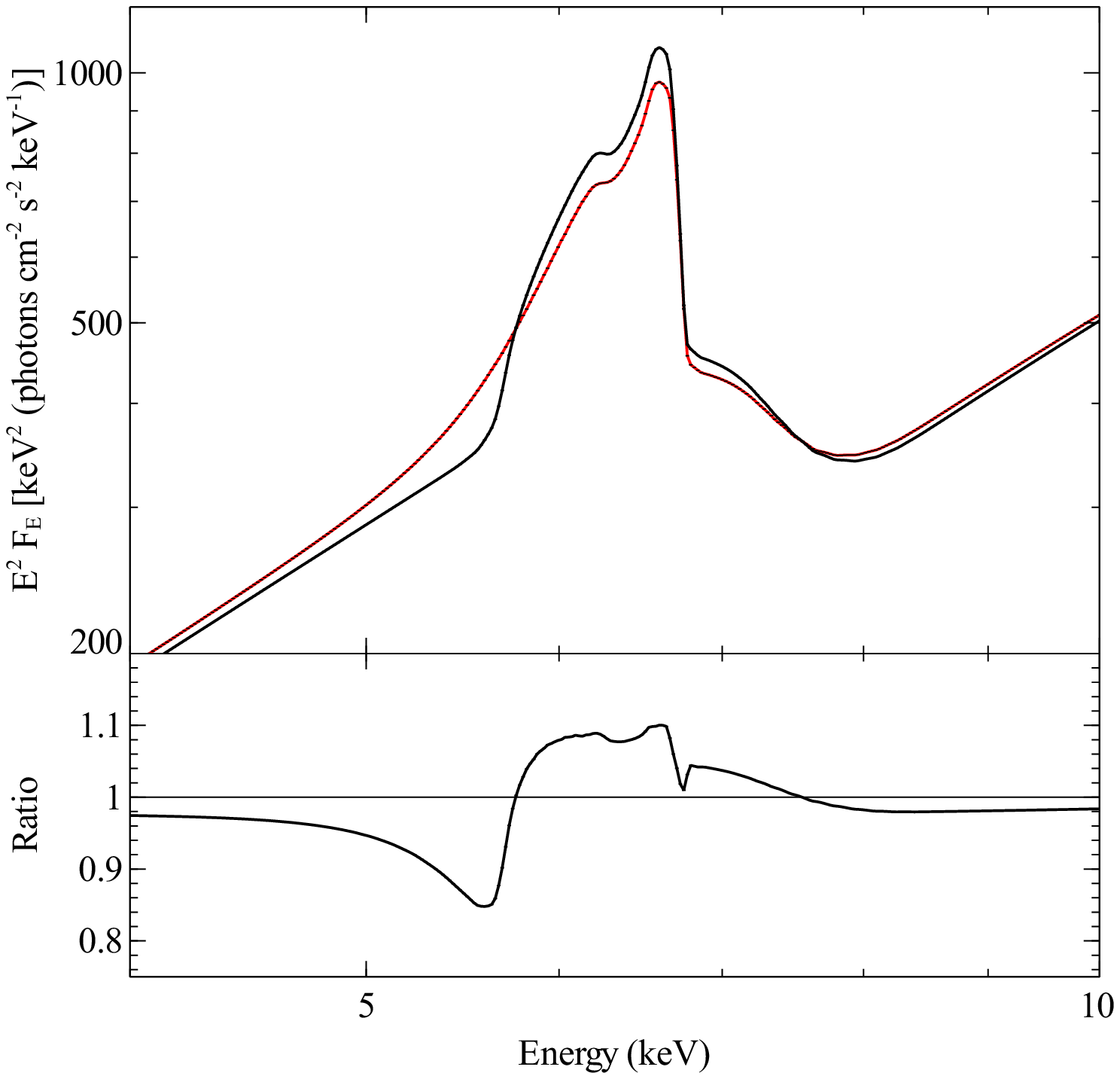}
  \caption{Upper panel: The solid line is the line profile for $h=30
    r_{\rm g}$ from Fig.~1. The dashed line shows the effect of
    truncating the disc at $r=30 r_{\rm g}$. Note that the most
    measurable change is relatively small and occurs below
    5.5~keV. Lower panel: As above but with reflection spectrum (Ross
    \& Fabian 2005) rather than just a line.  }
\end{figure}

For the purpose of illustration, we concentrate here on a single
emission line, but the principle is the same if the whole reflection
spectrum is used. We employ a simple lamp-post geometry for the
illumination, which means that the coronal emission source is
point-like, lies along the central rotation axis of the disc and emits
isotropically. We consider more complex geometries later.  We use {\sc
  relline-lp} of Dauser et al (2013) in the spectral-modelling package
{\sc xspec} to model the response of a single emission line from the
disc. This model is based on ray-tracing calculations, so accurately
represents the strong gravity effects of light bending and
gravitational redshift. The disc emissivity profile of the reflected
spectrum is implicitly included in {\sc relline-lp}. It can be
explicity specified using a broekn power-law with  {\sc relline},
its convolution kernel {\sc relconv}, or any other similar kernel.

\begin{figure}
  \centering
  \includegraphics[width=0.99\columnwidth,angle=0]{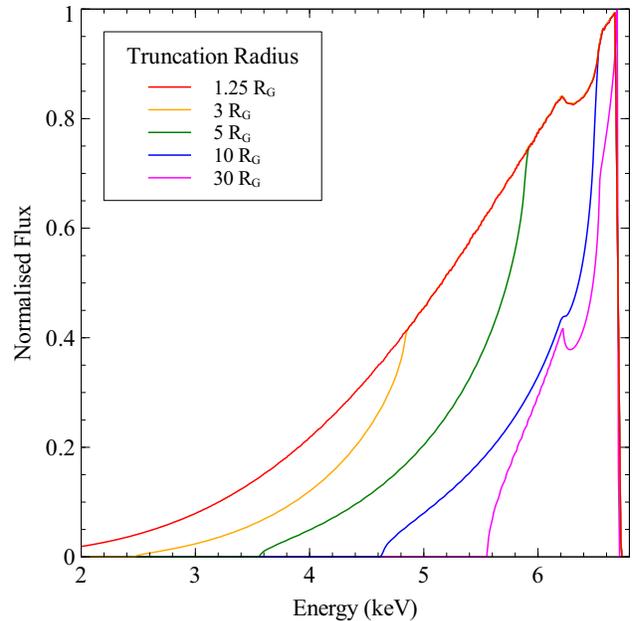}
  \caption{ Model line profiles predicted for a source height 
    $h=3r_{\rm g}$ above the centre of a disc truncated at $r_{\rm
      in}=1.25, 3, 5, 10$ and $30r{\rm g}$.     }
\end{figure}

The emission line predicted as the source is raised from 3 to 30
gravitational radii ($3-30 GM/c^2$) is shown in Fig.~1. The black hole
has a high spin of 0.998, and the disc continues down to the ISCO,
which is at $1.235r_{\rm _g}$. When the source is high above the disc,
at height $h>6r_{\rm _g}$, the line does not look very
broad. This is because little illumination is reaching the innermost
part near the ISCO. As the height is increased further the
illumination on the disc approaches the Newtonian pattern which is
roughly constant out to radius $r\sim h$, beyond which the emissivity
of the disc drops with radius as a power-law of index 3, as plotted in
Vaughan \& Fabian (2004) .

To highlight the effect of illumination on detection of the emission
from the ISCO when the source is at a height of $30r_{\rm g}$, we show
in Fig.~2 the line profile for $r=r_{\rm ISCO}=1.235 r_{\rm g}$ and
when the disc is truncated at $30r_{\rm g}$. The small difference in
normalization would in reality be unmeasurable since the absolute
value is unknown; the main indication of truncation lies below 5.5
keV, where the line strength is less than 10 per cent of its
peak. Truncation would be difficult to detect clearly with current
instruments and even a bright source.

The conclusion here is that at least part of the coronal source needs
to be situated at a height below at most $10r_{\rm g}$ in order that
clear evidence for truncation beyond the ISCO is obtainable. This also
applies to spin measurements, which only yield a lower limit to the
spin unless the coronal height is small so that the ISCO is well
illuminated (Dauser et al 2013). Any firm recommendation depends on
the quality of the data, and source conditions including iron
abundance, inclination and the spin itself, (Fig.~1 in Wilkins \&
Fabian 2011 shows an example of the energy range relevant to various
inner disc radii). The above recommendation, $h<10r_{\rm g}$, is
relevant to current data quality.

Note that similar results apply if the source is even a mildly
relativistic outflow, as suggested by Beloborodov (1999; see also
Malzac, Beloborodov \& Poutanen 2001). Due to
special relativistic aberration the emission would be beamed away from
the disc, mimicking the appearance of a high source. This could well
be relevant to the situation when a jet develops in a source where the
coronal emission occurs from the base of that jet (Markoff, Nowak \& Wilms
2005). The location of the base of that jet may be quasi-stationary, yet
the emitting matter within the base may flow rapidly upward. This
rapid motion is of course what is relevant for the emission pattern
and disc irradiation.

Both coronal height and motion could be relevant to jetted sources,
such as the low state of black hole binaries,  for example GX339-4, and
radio-loud AGN such as 3C120. Lohfink et al (2013) find that the inner
radius of the disc in 3C120 appears to change following a radio
outburst (see also Marscher et al 2002). Possible alternative interpretations
are that either coronal height has increased or the coronal material
has developed an outward velocity of a few tenths of the speed of
light.

When a disc is truncated and the inner radius is well illuminated then
the effects of truncation can be seen in the line profile (Fig.~3). In
this extreme  example the source height is only $3r_{\rm g}$. Note
that the peak of all model lines is normalised to unity and does not
show the drop in overall reflected emission which would result from the
larger truncation radii in this example.
 
\begin{figure}
  \centering
  \includegraphics[width=0.93\columnwidth,angle=0]{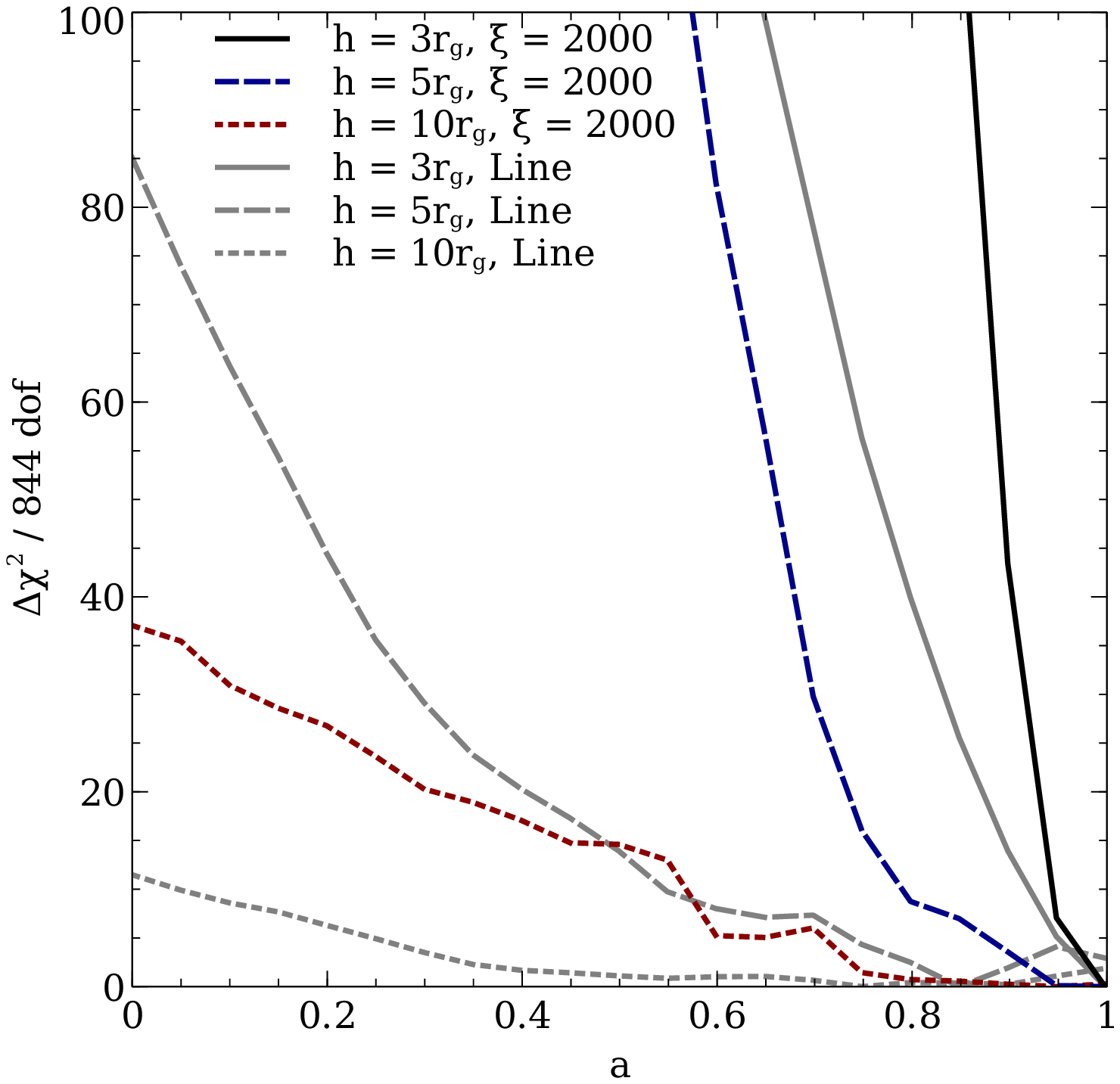}  
  \includegraphics[width=0.93\columnwidth,angle=0]{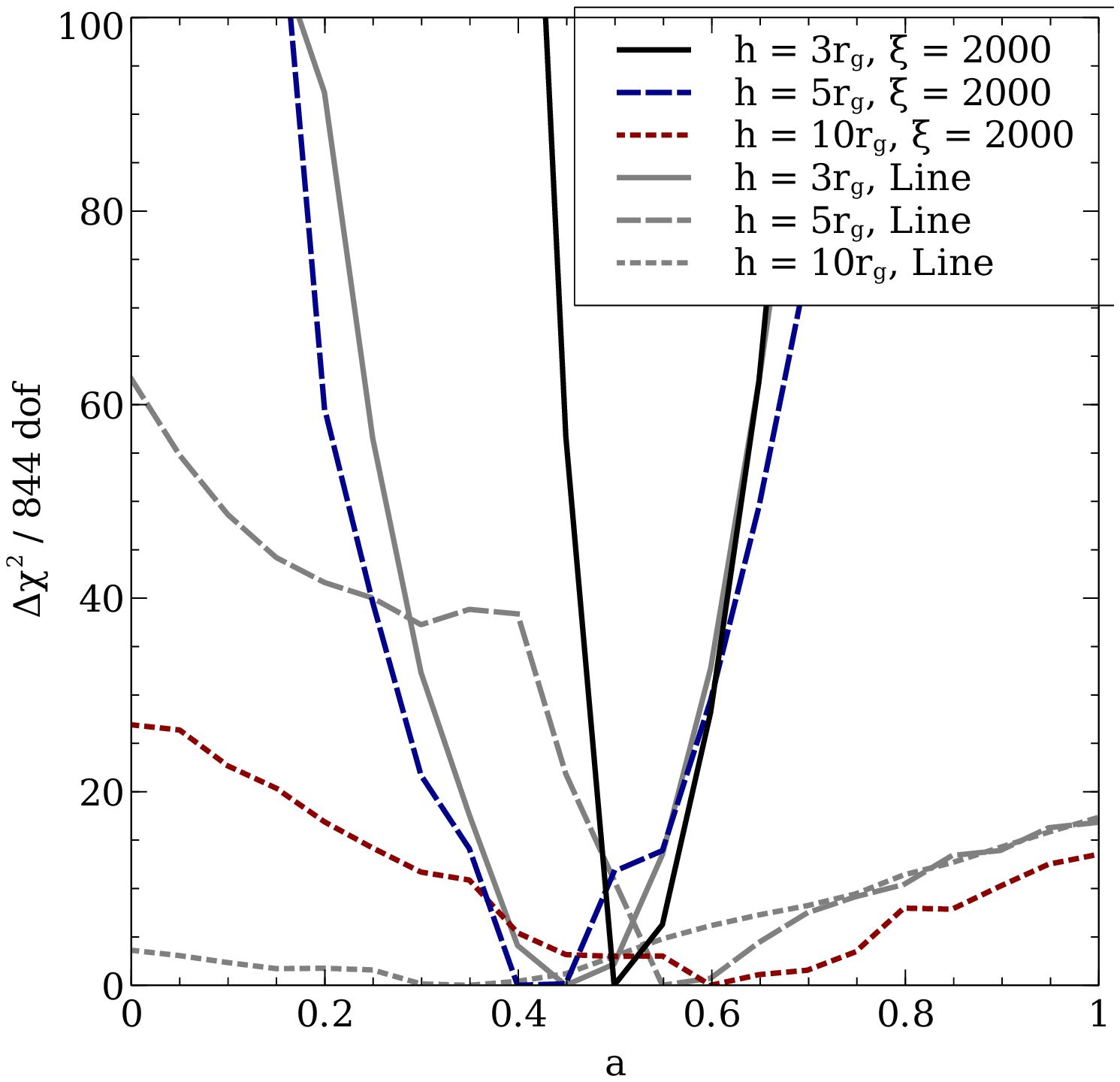}  
  \caption{Model fitting using the correct emissivity profile from
    {\sc relline-lp} to demonstrate the sensitivity to spin and disc
    ionization parameter. The highest sensitivity arises when the
    source height is low and thus best irradiates the ISCO and also if
    the disc is ionized. }
\end{figure}

\subsection{Simulations}

The height of the X-ray source above the disc is manifest in the
emissivity profile of the accretion disc used in the computation of
the broadened line profile. If the spin of the black hole is to be
obtained by fitting to the profile of the broadened emission line,
some assumption is required as to the emissivity profile of the disc
in the model that is fit to the data.

To quantify the effect of the assumed emissivity profile, spectra were
simulated using a model consisting of a power-law continuum and
relativistically-broadened iron K$\alpha$ line resulting from
illumination by an X-ray source at different heights ($3, 5 $ and
$10r_{\rm g}$) above the disc. (This extends the simulations shown in
Dauser et al 2013.) The emission line was modelled with the {\sc
  relline-lp} model and the fluxes of the continuum and reflected line
correspond to those found in the spectrum of the narrow line Seyfert 1
galaxy 1H0707-495 (Fabian et al 2009), but with a photon index
$\Gamma=3$ and disc inclination of 30 deg. A 500ks observation made with
XMM Newton was assumed using the {\sc fakeit} command in {\sc
  XSPEC}. In each case, the spin was set to be either 0.998
(Fig.~4. upper) or 0.5 (Fig.~4, lower). We also include examples where  
an ionized disc model (Ross \& Fabian 2005), with ionization
parameter $\xi=2000$, is assumed. There is now a deep smeared edge in
the spectrum which provides additional sensitivity and thus tighter
constraint on spin.

Just as for real observations, the spin is determined by fitting the
profile of the iron K$\alpha$ line in the fake spectra over the energy
range 0.3--10~keV, using either the {\sc relline-lp} model or {\sc
  relconv} convolved with {\sc reflionx} plus power-law continuum,
both of which assume the correct emissivity profile for the relevant
illumination case.

\begin{figure}
  \centering
  \includegraphics[width=0.85\columnwidth,angle=0]{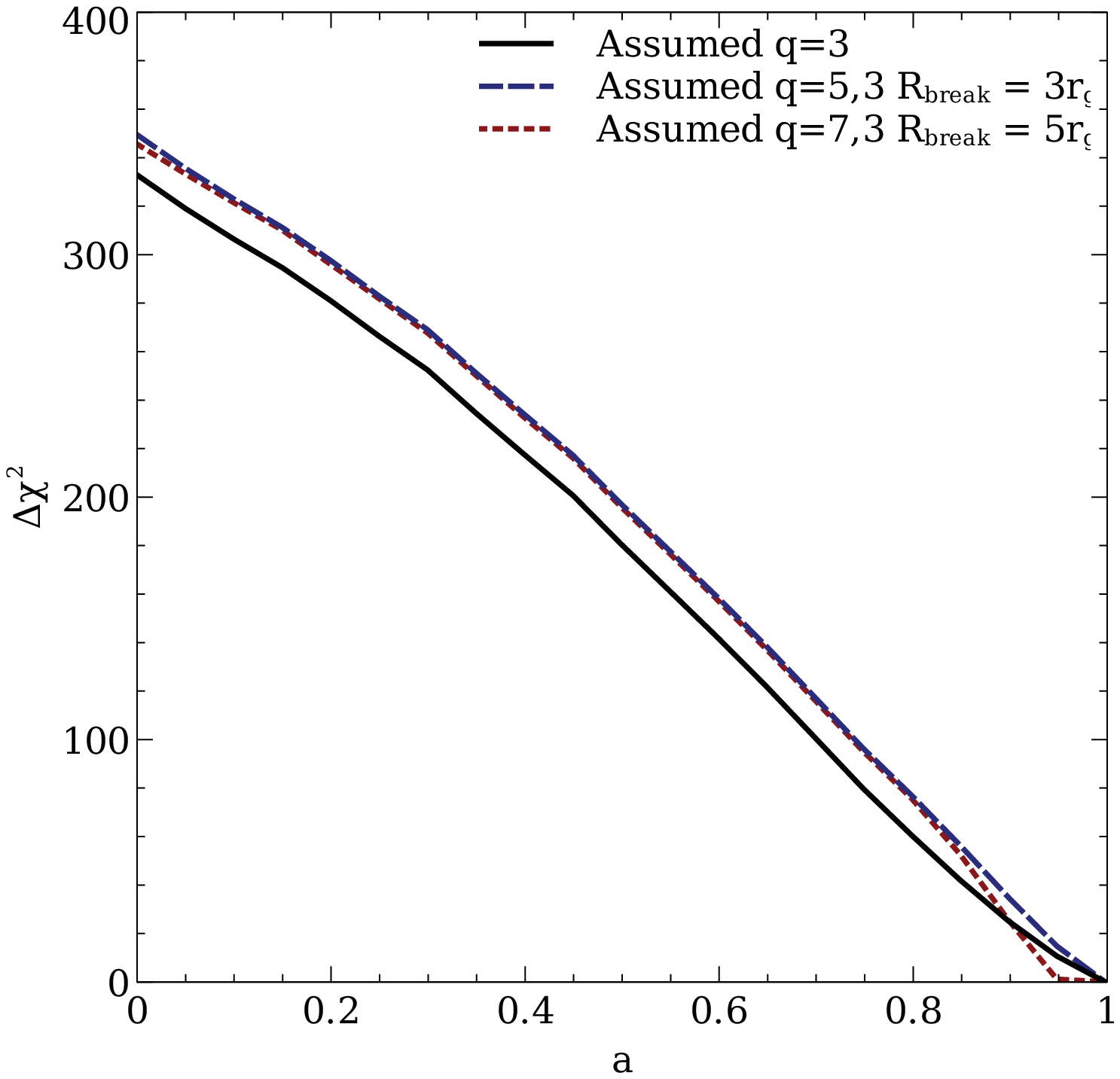}  
  \includegraphics[width=0.85\columnwidth,angle=0]{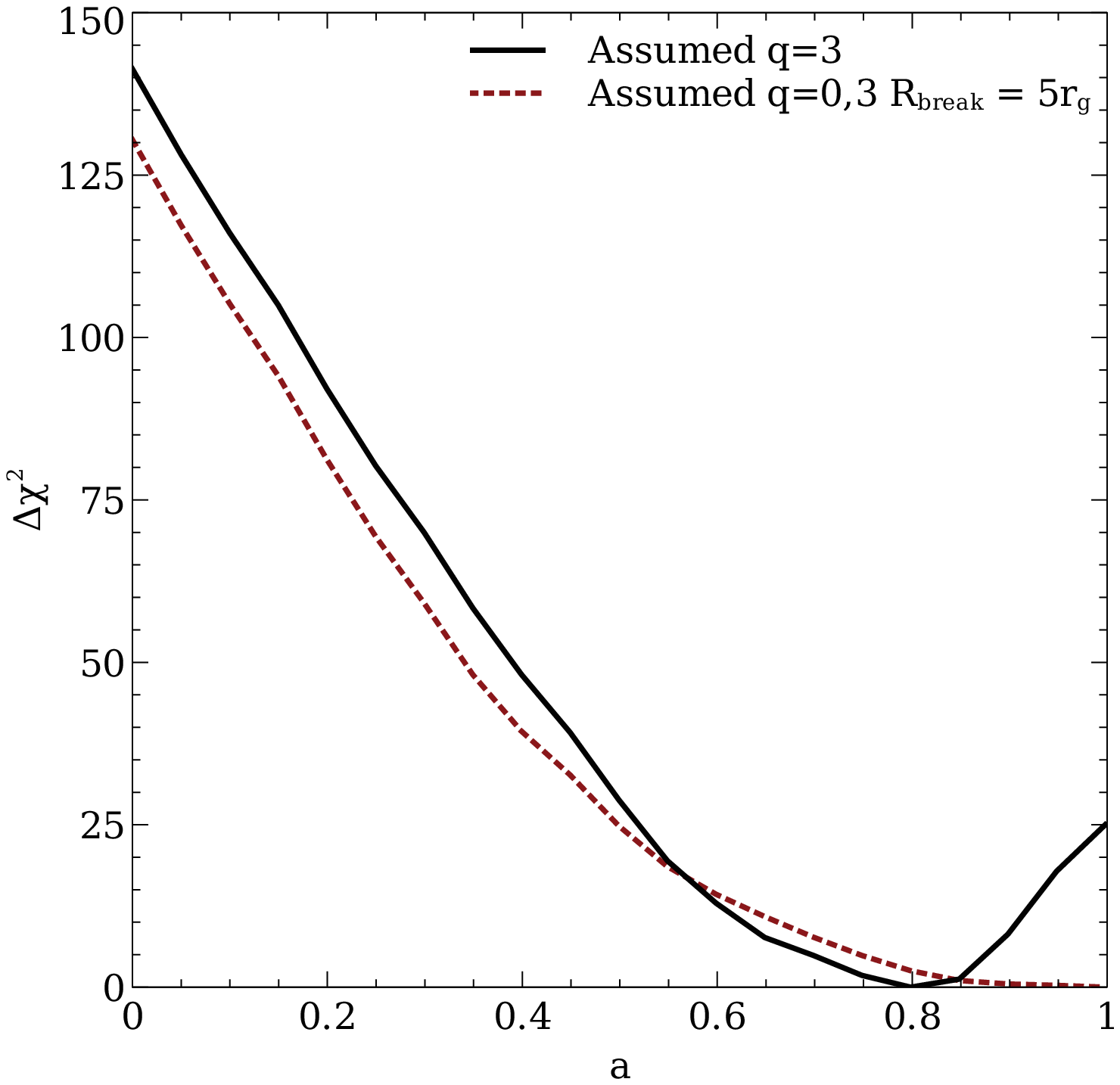}
  \includegraphics[width=0.85\columnwidth,angle=0]{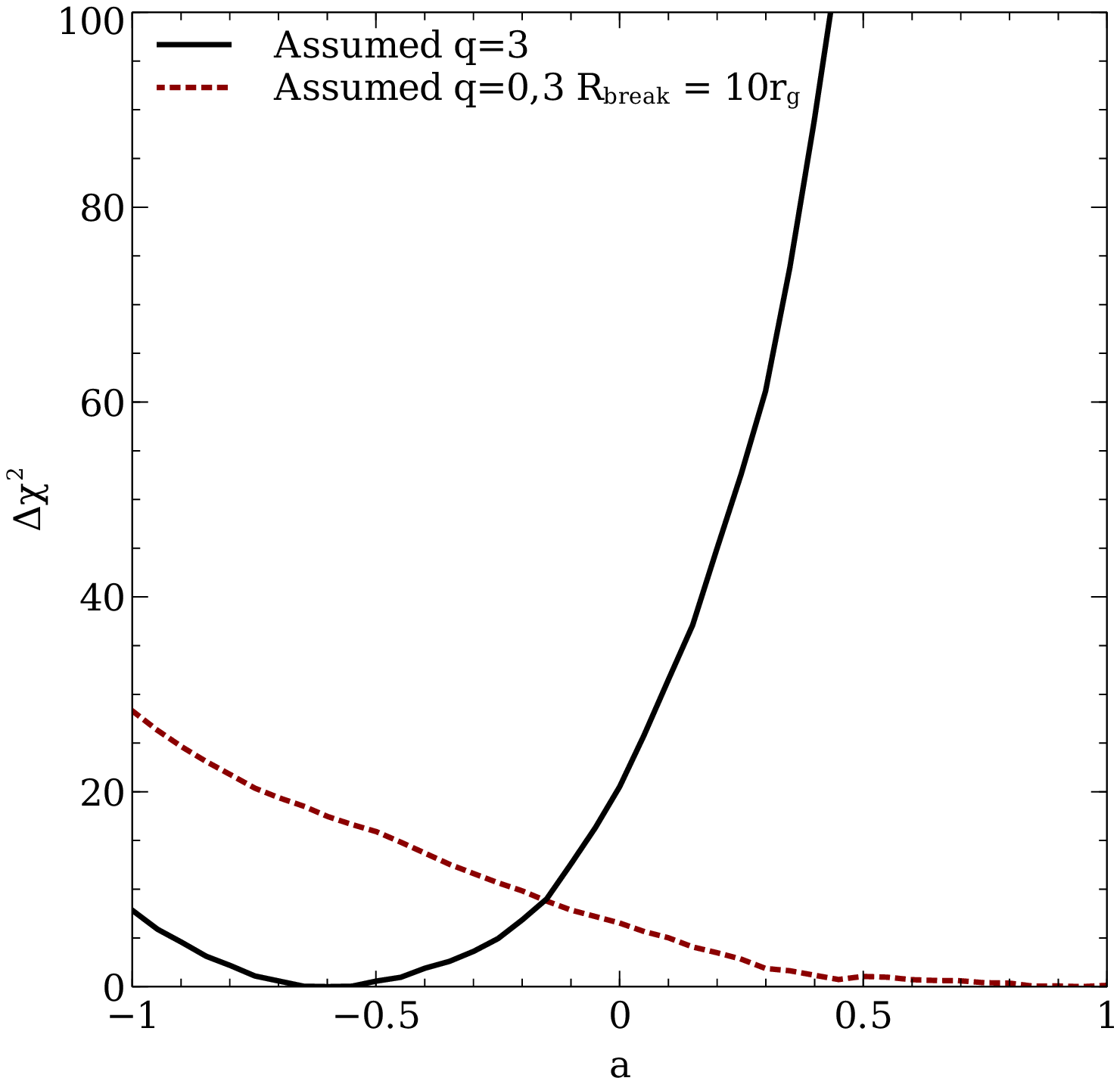}
  \caption{Model fitting using broken power-law approximations for the
    emissivity profile (see e.g. Wilkins \& Fabian 2012). Maximum spin
    is assumed with $r_{\rm in}=r_{\rm ISCO}$. The inner and outer
    power-law indexes, $q$, are listed together with the break
    radius. If only one index is given then a single power-law is
    assumed. Upper panel: $h=3r_{\rm g}$, Middle panel: $h=5r_{\rm
      g}$, Lower panel: $h=10r_{\rm g}$. Poor approximations lead to
    underestimation of spin. (The true spin in the simulation is
    0.998.)}
\end{figure}

In Fig.~5 we show results from simulations using {\sc relline}, which
uses power-law approximations of the emissivity profile.  In this
case, when the source is at a greater height above the disc,
overestimation of the emissivity index parameter (the power law slope)
causes the measured spin to be systematically underestimated
(Fig.~4). When the source is high, there exists a significant
flattened portion of the emissivity profile at radii between the
innermost peak and $r ~ h$ (Wilkins \& Fabian 2012). Even assuming the
Newtonian emissivity index $q=3$ causes the redshifted emission from
the inner part of the disc to be overestimated (since the overall
normalisation of the emission line is driven by the blueshifted
peak). The best fit to the observed spectrum therefore requires that
the disc be artificially truncated at larger radius to compensate for
this extra emission created from the inner part of the disc. This
artificial truncation of the accretion disc is interpreted as a lower
measurement of the black hole's spin.

Thus an assumed emissivity profile index of 3 can mean that the
spin is underestimated when the illuminating X-ray source is as little
as $5r_{\rm g}$  above the plane of the accretion disc, though it is
important to emphasise that this effect systematically underestimates
the black hole spin, meaning that conclusions drawn about rapidly
spinning black holes remain valid.

\section{More complex geometries}

The situation becomes more complicated if the coronal source is
extended, as it surely must be. Observations including reflection
spectra, reverberation and microlensing (Fabian et al 2012; Kara et al
2013; Morgan et al 2012) indicate that the corona in at least some sources
is compact, meaning confined to $r\sim 10 r_{\rm g}$ or less (Fabian
2012; Reis \& Miller 2013). This seems plausible for rapidly spinning
sources where most of the disc energy release occurs within that
radius\footnote{Half the disc power is emitted within $5 r_{\rm g}$ for a
  highly spinning black hole and within $\sim33 r_{\rm g}$ at low
  spin, Thorne (1974). If $h\sim r$ in radio-quiet objects, then
  $h\sim 5+28(1-a)$.}.  The corona is presumably powered by magnetic
fields anchored in the disc so any quasi-stationary corona is likely
to have a height $h< 10-30 r_{\rm g}$, with smaller values applying to
higher spin.

Even in such cases, the effects of the inevitable strong light bending
mean that the source appears anisotropic to the observer. This is the
essence of the light-bending model (Miniutti \& Fabian 2004) where the
intensity of the reflection spectrum changes with source height
relatively little whereas the observed primary coronal emission
changes markedly. 

If the source is extended uniformly in height then the lower regions,
which give the least direct emission, will give the most irradiation
to the innermost parts of the disc near the ISCO, whereas the opposite
applies for the upper parts. If the Comptonization spectrum of the
corona changes with height (the electron temperature and optical depth
are unlikely to be constant with height) then it is possible that the
innermost parts near the ISCO are irradiated by a power-law of
different photon index $\Gamma$ to that which characterizes the mean
observed power-law spectrum. Similar effects should occur for a
radially-extended source.

They do not necessarily make spin difficult to measure\footnote{On the
  basis of the emissivity profile, Wilkins \& Fabian (2012) show that
  the corona in 1H0707-495 is extended. There is however significant
  reflection from the inner disc so  the spin measurement is robust.}
but do mean that the spectral models may need to be more sophisticated
than those commonly adopted in which $\Gamma$ for the incident
power-law is usually forced to be the same as that for the reflection
component. The observed spectral turnover in the power-law continuum
may also be different from that deduced from the reflection spectrum.

Propagating spectral variations produced as power flows through a
spatially-extended corona, can lead to low-frequency spectral lags as
discussed by Kotov et al (2001) and Wilkins \& Fabian (2013). The
common occurrence of such lags (Miyamoto \& Kitamoto 1989; Nowak et al
1999; Gilfanov et al 1999; Poutanen 2001; McHardy et al 2007) is good
evidence that coronae are extended and inhomogeneous.

The inner parts of extended coronae could produce a softer spectrum
(higher $\Gamma$) than the outer parts. The energy density in soft
disk photons which are upscattered by the hot electrons to make the
coronal continuum should peak within a few $r_{\rm g}$ for rapidly
spinning black holes. Its ratio to the energy density of the electrons
is an important factor in determining $\Gamma$. Consequently,
estimates of spin from the soft excess in the reflection component, such as
presented by Walton et al (2013), could give higher spin values for
the same object when compared with those obtained from iron-K
measurements. We assume that the simplest model is used, where the
reflection and power-law components have the same $\Gamma$ and a
single source height, i.e. emissivity profile. IF the inner regions of
the corona produce a harder spectrum then of course the opposite
conclusion could apply.

\section{Systematics in the X-ray reflection method}

We now outline some of the systematics inherent in the measurement of
spin and inner radius using the X-ray reflection method. Beside the
level at which the inner radii are illuminated, relevant issues
include the ionization, density, elemental abundances, level of
surface turbulence and uniformity of the disc, the extent to which the
corona is extended and thus whether a correct emissivity model is
adopted. The important issue of whether the ISCO and $r_{\rm in}$
coincide has already been discussed in the Introduction.

The ionization state of the surface of the disc, which determines the
ionization parameter $\xi$ in the reflection model (e.g. either {\sc
  reflionx} (Ross \& Fabian 2005) or {\sc xillver} (Garcia \& Kallman
2010); Garcia et al 2013a), affects the overall shape of the reflection
spectrum. If the ionization parameter is high, the (vertically)
outermost disc gas
is highly ionized and much of the iron line scatters out from below
the outer Thomson depth, leading to additional line broadening due to
Compton scattering (Ross \& Fabian 2007). The ionization state depends
upon the illumination level and pattern (i.e. radial dependence) and
on the disc density profile. If, as expected, the power for the corona
is extracted magnetically from the disc, then it is not obvious that
the radial dependence of that extraction should match that of the
analytic dissipation relation of a Shakura-Sunyaev (1973) or
Novikov-Thorne (1974) disc, nor, therefore, that the density profile
is known or can be found other than by observation. Svoboda et al
(2012) show some effects produced by varying the ionization parameter
across the disc.  An extended corona exacerbates the uncertainties in
the ionization profile.

Svoboda et al (2009) discuss the effects produced by limb darkening,
or brightening, of the emergent reflected radiation. Note that the
vertical structure of the disk surface depends on the pressures
involved, which in most cases are dominated by magnetic fields (Blaes
2013), so simple thermal hydrostatic support has little
relevance. There are currently no detailed simulations of irradiated
discs to inform us of the expected density structure. Commonly used
constant density models are an approximation.  The vertical density
profile of the disc surface affects the outer ionization parameter. A
decreased density leads to an increased ionization which can affect
the degree of limb darkening.

Disc inclination leads to further uncertainties. In principle it is
measurable by the blue wing of the line, which is sensitive to the
increased doppler shifts produced by higher inclination. Measurements
of spin and inclination are correlated, but in a model dependent
manner; for examples see Tomsick at al (2014). In addition, there are
intrinsic inclination effects on the reflection spectrum, now modelled
by {\sc relxill} (Garcia et al 2013b).

Numerical models of the inner regions of thin discs show them to be
turbulent (Armitage \& Reynolds 2003; Beckwith et al 2008). This could
lead to the surface density of the disc being patchy (see e.g. Dexter
\& Quataert 2012), thus leading to a range of ionization parameter at
each radius. There might moreover be additional doppler broadening due
to the turbulent motions (Tomsick et al 2014). This is presumably
subsonic, which means that it should not be important for objects
where a low disc ionization, and thus low disc surface temperature, is
inferred.

Finally we note that the geometry of the corona is probably not known
{\it a priori} so a model-independent approach is required for
determining the emissivity profile. Singly or doubly broken power-law
models are a good guide (Wilkins \& Fabian 2012).

\section{Selection effects}

Observational selection effects mean that the brightest sources in the
Sky will tend to have both high spin and inclination, even if the
underlying distribution favours no spin or inclination. If the mass
accretion rate is independent of black hole spin, then the increase in
efficiency of energy release with spin means that the most rapidly
spinning black holes will be the most luminous and thus dominate
flux-limited samples (Brenneman et al 2011; Reynolds et al
2012). Similarly, if the corona corotates with the black hole or inner
disc, then special relativistic aberration will make the source
brightness increase with inclination; the same will be true of
reflection or intrinsic disc emission (Sun \& Malkan 1991; Walton et
al 2013). A puffed-up disc, an equatorial accretion flow (or outflow),
a warped disc or a surrounding torus may however prevent sources at
very high inclinations from being prevalent.

Note that the degree to which the corona is corotating above the inner
disc of a rapidly spin black hole will affect the apparent reflection
fraction. This provides in principle a method for determining the
rotation rate of the corona.

\section{Discussion}

Robust measurements of black hole spin and disc
truncation from X-ray reflection in accreting black holes using  
current data require that at least part of the coronal emission source
lies below about $10r_{\rm g}$. This is necessary in order that the
ISCO is illuminated well. Narrowing of the iron line could be
associated with disc truncation, coronal elevation or mild
relativistic outflow of the coronal source.  Since jetted radio
emission is commonly seen in the low state of stellar mass black
holes, it is plausible that either of the last two effects, or their
combination, could be responsible and their effect needs to be
eliminated before disc truncation can be reliably deduced. Independent
evidence for disc truncation, from thermal disc emission for example, is
not affected by the height of the corona. 

Subject to the uncertainty in the relation between the ISCO and
$r_{\rm in}$, excellent estimates of spin and inner disc radius are being
made with  X-ray reflection using the methods outlined above. The
robust measurements of spin from the reflection method which have
yielded high spin are perfectly valid and simply indicate that much of
the irradiating radiation originates from within $10r_{\rm g}$. The
effects discussed above suggest that less certain results (due to the 
corona being higher or outflowing, for example) indicate lower limits
on spin.

\section*{Acknowledgements} We thank the referee, Javier Garcia, for
helpful comments.

%\bibliographystyle{mnras}
%\bibliography{refs}

\end{document}